\documentclass{jetpl}
\twocolumn
\lat

\usepackage{amsfonts}
\usepackage{amsmath}
\usepackage{amssymb}

\usepackage[T1]{fontenc}

\usepackage[dvips]{graphicx}
\usepackage{myaasmacros}
\usepackage{color}
\usepackage{multirow}

\def\bea{\begin{eqnarray}}
\def\ena{\end{eqnarray}}

\newcommand{\bn}{\mathbf{n}}

\title{Full sky harmonic analysis hints at large UHECR deflections}

\rtitle{Full sky harmonic analysis hints at large UHECR deflections}
\sodtitle{Full sky harmonic analysis hints at large UHECR deflections}

\author{P. G. Tinyakov\thanks{E-mail: petr.tiniakov@ulb.ac.be}, F. R. Urban\thanks{E-mail: furban@ulb.ac.be}}

\address{Universit\'e Libre de Bruxelles, Service de Physique Th\'eorique, CP225, 1050, Brussels, Belgium}

\dates{}{*}

\abstract{The full-sky multipole coefficients of the ultra-high energy cosmic ray (UHECR)
flux have been measured for the first time by the Pierre Auger and Telescope
Array collaborations using a joint data set with $E>10$~EeV. We calculate
these harmonic coefficients in the model where UHECR are protons and sources
trace the local matter distribution, and compare our results with
observations. We find that the expected power for low multipoles (dipole
and quadrupole, in particular) is sytematically higher than in the data: the
observed flux is too isotropic. We then investigate to which degree our
predictions are influenced by UHECR deflections in the regular Galactic
magnetic field (GMF). It turns out that the UHECR power spectrum coefficients
$C_\ell$ are quite insensitive to the effects of the GMF, so it is unlikely
that the discordance can be reconciled by tuning the GMF model. On the
contrary, a sizeable fraction of uniformly distributed flux (representing for instance an
admixture of heavy nuclei with considerably larger deflections)
can bring simulations and observations to an accord.}

\begin{document}

\maketitle

\label{firstpage}

\section{Introduction}
\label{sec:intro}

Despite the fact that the actual sources of UHECRs have still not been
identified, it is rather natural to expect them to follow, to some
extent, the large scale structure (LSS) observed in the sky. Indeed,
the propagation distance of UHECRs of above $10^{19}$~eV is limited to
several hundred Mpc due to their interaction with the inter-galactic medium (IGM) \cite{Greisen:1966jv,Zatsepin:1966jv}. The matter
distribution is not homogeneous over such distances, hence if UHECRs
are extragalactic, one expects an anisotropy in their arrival
direction distribution, reflecting the inhomogeneity of the source
distribution.  Such anisotropies, on a sphere, can be revealed via a
harmonic analysis, where the coefficients of the complete set of
spherical harmonics carry the information, multipole by multipole,
about the possibly non-uniform UHECR flux. The harmonic analysis is
thus a way to compress the data in a form most suitable for
statistical tests.

Recently the Telescope Array (TA) and Pierre Auger Observatory (PAO)
collaborations have joined forces to provide the first full-sky ultra-high
energy cosmic rays (UHECRs) map~\cite{Aab:2014ila}.  With single earth-based
experiments being forcedly blind to a big chunk of the sky, only the combined
data sets from two --- or more --- machines can provide the complete
picture. This is particularly important for the harmonic analysis which, as
detailed in the joint TA/PAO paper, strongly benefits from a whole sky
coverage, both theoretically/qualitatively (no need to assume anything about
the flux) and practically/quantitatively (some errors are significantly
suppressed)~\cite{Sommers:2000us}. Joining the data of the two experiments thus
made possible, for the first time, to measure the harmonic multipoles of the
UHECR flux distribution in an assumption-free way.

One natural question is then: is the harmonic power spectrum expected
from the LSS the same as that actually observed? 
The caveat here is that UHECRs do not travel on a straight line from
source to the Earth, because of the magnetic fields (MFs) they
encounter on their way; these deflect their trajectories and mask the
original arrival directions, and with that the sources or UHECRs. The
most relevant MF in this respect is housed by our own Galaxy (GMF),
with strength in the $\mu$G range, see for instance~\cite{Haverkorn:2014jka} and references therein.  The GMF is separated into
large- (regular or coherent) and small- (turbulent or random) scale
components, the regular part dominating the CR deflections. When
combined, these fields are expected to steer 10 EeV protons by a
few degrees, far away from the galactic plane, to up to 
several tens of degrees at very low galactic latitudes. Now, what
does this mean for the harmonic analysis?  Are the anisotropies
erased, or is the power spectrum distorted?

What we find in this analysis is that there is a striking
mismatch between the power spectrum we simulate from the LSS,
\emph{assuming a purely protonic primary composition}, and the one
reconstructed from the data: the amplitudes of low multipoles
(particularly, the quadrupole $C_2$, the second momentum in the
harmonic decomposition) in the data are significantly lower than the
calculated LSS ones. The scope of this work is to delve deeper into
this issue; in particular, we want to understand the r\^ole of the GMF
in this result.

Before embarking on the analysis, we briefly summarise our findings:
\begin{itemize}
  \item there is a lack of power in the low multipoles (notably, dipole
    and quadrupole) as observed
    by TA/PAO compared to the expectations from protons tracing LSS;
  \item the regular GMF shuffles direction-dependent single harmonic
    coefficients, demonstrating how these are not fully reliable indicators
    of source anisotropy;
  \item however, the power spectrum is barely affected by the regular
    GMF, which means that the latter can not bring observations and
    simulations to an accord;
  \item the random GMF has also very little effect on the low multipoles;
  \item a moderate fraction of uniformly-distributed events (which could,
    for instance, represent an of heavy nuclei) instead does temper the
    tension between data and expectations, for it contributes to the
    isotropisation of the signal even on largest angular scales.
\end{itemize}

For the rest of the paper we will begin summarising the results of the joint
TA/PAO analysis in Section~\ref{sec:joint}~; then we will introduce the
simulated power spectra from the LSS, and discuss the missing quadrupole
problem in Section~\ref{sec:LSS}~.  The impact of the GMF on this result is
detailed in Sec.~\ref{sec:GMF}~, whereas the turbulent GMF is discussed in
Sec.~\ref{sec:Fe}~, alongside the effect on the power spectrum of a different
composition of cosmic rays primaries.  We will conclude in
Section~\ref{sec:end}~.

\section{The joint TA/PAO analysis}
\label{sec:joint}

As any angular distribution on the unit sphere, the flux of cosmic ray
$\Phi(\bn)$ in a given direction $\bn$ can be decomposed in terms of a
multipolar expansion onto the spherical harmonics $Y_{\ell m}(\bn)$:
\begin{equation}
\label{eq:ylm}
\Phi(\bn)=\sum_{\ell\geq0}\sum_{m=-\ell}^\ell a_{\ell m}Y_{\ell m}(\bn) \, .
\end{equation}
Anisotropy fingerprints are encoded in the $a_{\ell m}$ multipoles.
Non-zero amplitudes in the $\ell$ modes contribute in variations of
the flux on an angular scales of about $\pi/\ell$ radians.

Cosmic ray events, in this language, are then simply sample points
for the underlying sources distribution on the sphere.  However, because the
sky coverage is non-uniform, what these events are sampling is the 
flux times exposure distribution.  Now, with full-sky but
non-uniform coverage, the customary recipe~\cite{Aab:2014ila} for decoupling directional
exposure effects from anisotropy ones consists in weighting the
observed angular distribution by the inverse of the \emph{relative}
directional exposure function $\overline{\omega}_r(\mathbf{n})$, so
that, inverting Eq.~(\ref{eq:ylm}), the actual data points are
unbiassed estimators of the underlying flux:
\begin{equation}
\label{eq:alm}
\hat{a}_{\ell m}=\sum_{i=1}^N\frac{Y_{\ell m}(\bn_i)}{\overline{\omega}_r(\bn_i)} \, ,
\end{equation}
where one can prove~\cite{Aab:2014ila} that upon averaging over a large number of 
realisations one has $\left\langle\hat{a}_{\ell
 m}\right\rangle= a_{\ell m}$.  Here $N$ is the number of events,
which are described as Dirac delta functions centred at the actual
arrival directions $\bn_i$.

While the individual $a_{\ell m}$ coefficients are
direction-dependent, the angular power spectrum coefficients $C_\ell$,
defined as averages of $|a_{\ell m}|^2$ over $m$,
\begin{equation}
\label{eq:cell}
C_\ell=\frac{1}{2\ell+1}\sum_{m=-\ell}^\ell \left|a_{\ell m}\right|^2 \, ,
\end{equation}
are rotation-independent quantities. Given that the regular GMF 
results in a (direction-dependent) rotation of the events, one
might expect that the power spectrum coefficients $C_\ell$ are much
less sensitive to the presence of GMF than individual amplitudes
$a_{\ell m}$. 

Now, in order to achieve full-sky coverage the data of two
different experiments must be combined; hence, the total exposure has to be
cross-calibrated in order to not introduce spurious effects in
Eq.~(\ref{eq:alm}). The details of the cross-calibration procedure,
and its performances, do not matter for us here, but can be found in
Ref.~\cite{Aab:2014ila} (see also~\cite{Array:2013dra}, \S2). Note, however, that in the cross-calibration
procedure only a small subset of all events -- those in the region of
overlapping exposures --- are used, and that the cross-calibration
errors propagate mainly into the $m=0$ components of the coefficients
$a_{\ell m}$ (in equatorial coordinates).

The data sets used in the analysis consist of UHECRs with energies
above 10~EeV, which amounts to 8259 for PAO, and 2130 for TA.
Table~\ref{tab:alm} reports the results for the $a_{\ell m}$
coefficients as presented in the TA/PAO joint paper.
\begin{center}
\begin{table*}
\begin{minipage}{\textwidth}
\begin{center}
\begin{tabular}{c|c|c||c|c|c||c|c|c}
  $\ell$ & $m$ & $a_{\ell m}$ & $\ell$ & $m$ & $a_{\ell m}$ & $\ell$ & $m$ & $a_{\ell m}$ \\
  \hline
  \hline
	&	&			&	&	& 			&	& -3	& -0.022 $\pm$ 0.034 \\
	&	&			&	& -2	& 0.038 $\pm$ 0.035	&	& -2	& 0.030 $\pm$ 0.039 \\
	& -1	& -0.102 $\pm$ 0.036	&	& -1	& 0.067 $\pm$ 0.040	&	& -1	& 0.067 $\pm$ 0.037 \\
  1	&  0	& 0.006 $\pm$ 0.074	& 2	&  0	& 0.017 $\pm$ 0.042	& 3	&  0	& -0.027 $\pm$ 0.040 \\
	&  1	& -0.001 $\pm$ 0.036	&	&  1	& 0.004 $\pm$ 0.040	&	&  1	& 0.009 $\pm$ 0.037 \\
	&	&			&	&  2	& 0.040 $\pm$ 0.035	&	&  2	& -0.004 $\pm$ 0.039 \\
	&	&			&	&	&			&	&  3	& -0.011 $\pm$ 0.034 \\
  \hline
  \hline
  \multicolumn{3}{|c||}{$C_1= 0.0035 \pm 0.0024$} & \multicolumn{3}{c||}{$C_2 = 0.0016 \pm 0.0014$} & \multicolumn{3}{c|}{$C_3 = 0.00097 \pm 0.00088$} \\
  \hline
\end{tabular}
\vspace{5pt}
\caption{Dipole, quadrupole, and octupole moments, with their uncertainties, in equatorial coordinates.  These are normalised so that the $a_{\ell m}$ measure the relative deviation with respect to the monopole $a_{00}$, that is, the $a_{\ell m}$ are redefined such that $a_{\ell m} \rightarrow \sqrt{4\pi}a_{\ell m}/a_{00}$.}
\label{tab:alm}
\end{center}
\end{minipage}
\end{table*}
\end{center}

As one may see, there are no statistically significant deviations from
isotropic expectation in any of the harmonic coefficients, the largest
discrepancy being in the value of $a_{1,-1}$. One also observes
that the errors are systematically larger for $m=0$, particularly for
the dipole $\ell =1$: a consequence of the cross-calibration
procedure.

\section{Large scale structures and anisotropies}
\label{sec:LSS}

In order to compare the power spectrum reconstructed from the data with the
expectations from the LSS model, we need to build the flux map which we are
going to sample with random Monte Carlo events and derive expectations for the
multipole coefficients. The procedure we used to build the expected flux is
described in detail in~\cite{Koers:2009pd,AbuZayyad:2012hv}. We first choose a
galaxy catalogue, in this case the 2MASS Galaxy Redshift Catalog (XSCz)
that is derived from the 2MASS Extended Source Catalog (XSC). The flux is
calculated from the flux-limited subsample of galaxies with the apparent
magnitude $m<12.5$ at distances $D<250$~Mpc by the method described in
\cite{Koers:2009pd,Koers:2008ba,Abbasi:2010xt}. The contribution from beyond 250~Mpc is considered uniform. All
galaxies are assumed to have the same intrinsic luminosity in UHECR. To
determine their individual contributions to the total flux we propagate
protons to the Earth taking full account of the redshift, distance and
attenuation effects. Individual fluxes are then smeared with a Gaussian
distribution of an angular width $\theta$, which is a free parameter. This is
done to account for limited detector resolution, and, most importantly, the
effects of the regular and turbulent GMF.  In this Section we will not attempt
to reconstruct the original direction of the events through the coherent GMF,
which is instead investigated in detail in Sec.~\ref{sec:GMF}.  Finally, where
this is relevant, the flux map is weighted accounting for the non-uniform
exposure of the actual experiment (or experiments).

\begin{figure}
\begin{center}
  \includegraphics[width=0.48\textwidth]{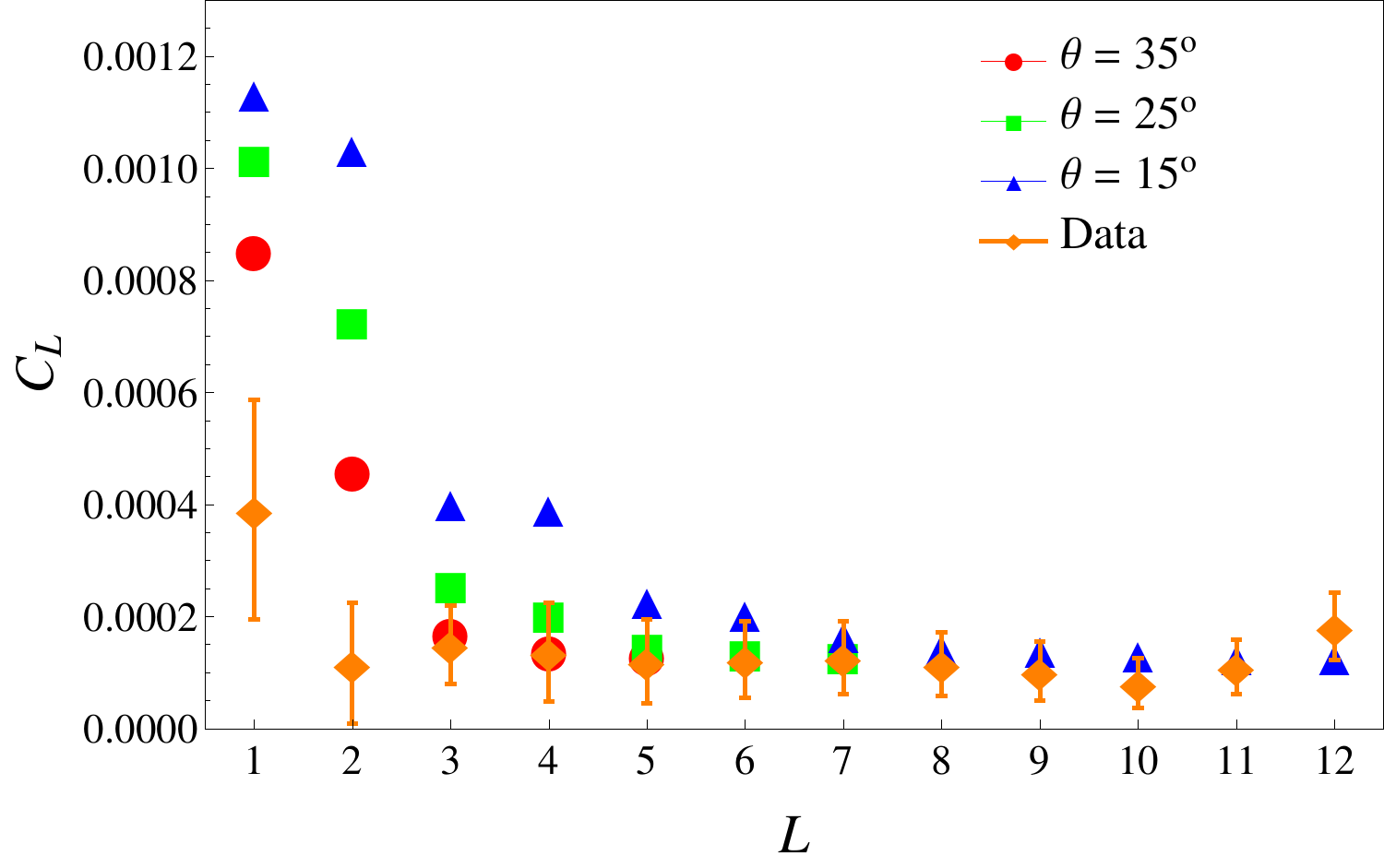}
\end{center}
\caption{The UHECRs angular power spectrum from the data and from
  simulated flux maps at different smearing angles, see the main text
  for details.}
\label{fig:lss}
\end{figure}

With the map of the expected flux on Earth at hand we simulate random sets of
cosmic ray events that this flux distribution would produce. Each mock set has
the same number of events as the actual data. We then calculate the harmonic
coefficients and the power spectrum for each of these mock sets. For each
harmonic coefficient we determine the mean value and the variance; we generate
as many mock sets as is necessary to make the variances negligible. In
Fig.~\ref{fig:lss} we show the result of this procedure: orange diamonds are
the actual data points with their errors; blue triangles, green boxes, and red
circles are the expectations from the simulations with smearing angles of
$\theta = 15^\circ,\;25^\circ,\;\text{and }35^\circ$,
respectively\footnote{Smearing the flux with a given $\theta$ by definition
  wipes away any power for multipoles $\ell \gtrsim \pi/\theta$, so we do
  not include multipoles for which by construction there is no power.}.

The most striking feature of this plot is the considerable tension between the
power in the low multipoles (notably, the dipole and quadrupole) expected from
the galaxy distribution, and what is observed in the data, the predicted power
being systematically higher. With a smearing angle of $15^\circ$, both the
dipole and quadrupole components of the flux are expected to vary at the level
of $\sim 10\%$, while no flux variations are detected in the data. However,
because the dipole measurement has larger error, the discrepancy is most
prominent in the case of the quadrupole. As the smearing angle grows the 
expected flux variation is watered down, the larger the multipole number, the
faster the isotropy sets in. 

All the higher multipoles are more or less within their expected values in the
LSS case, although at the level of precision currently attainable with TA and
PAO these are difficult to distinguish from simple isotropy. We will devote
the rest of the paper to discuss this observation, and to
understand and clarify the r\^ole of the GMF in drawing conclusions from it.

A comment is in order at this point. As we have seen, the measurement of the
power spectrum coefficients $C_\ell$, in particular $C_1$, is
obscured  by the cross-calibration procedure which introduces a
large error. One may define observables that are free from this problem (note
that, as we will argue in the next section, such observables would also lose 
an important advantage of $C_\ell$: their insensitivity to the regular magnetic
field). These are the coefficients $c_n$ of the Fourier decomposition of the flux
in right ascension $\phi$ defined as follows,
\[
  c_n \equiv \frac12 \int\text{d}\bn \Phi(\bn) Y_n(\phi) \, ,
\]
with $Y_n(\phi) \equiv (\sqrt2\cos{n}\phi, 1, \sqrt2\sin|n|\phi) /
\sqrt{2\pi}$ for $n>0$, $n=0$, and $n<0$, respectively.  The coefficients 
$c_n$ can be measured in a single experiment without making any assumption about 
the flux.  They are obviously free from the errors introduced by the 
cross-calibration procedure\footnote{Although it is indeed possible to 
determine these coefficients univocally from a single experiments, it is not 
guaranteed that in the north and south emisphere they will agree with each 
other.}.

In order to make contact with previously defined quantities, we note that
$c_n$ can be expressed in terms of spherical harmonic coefficients $a_{\ell
  m}$ with $l\geq |n|$ (see the appendix for a derivation). So we may use the
harmonic coefficients $a_{\ell m}$ calculated above for the LSS model to infer
the LSS prediction for $c_n$. The two lowest coefficients $c_{\pm 1}$ receive
contributions from the projection of the dipole and of odd $\ell$ higher 
multipoles onto the right ascension plane. For the LSS model the contribution 
of the dipole is dominant, and we may approximately write
\begin{eqnarray}
\nonumber
c_{+1} &\simeq& \frac{\sqrt3\pi}{8} a_{1,-1} \, ,\\
\label{eq:c_+-}
c_{-1} &\simeq & \frac{\sqrt3\pi}{8} a_{1,1} \, .
\end{eqnarray}

In order to compare to the measurements, the coefficients (\ref{eq:c_+-})
can be combined into an amplitude $d^2 \equiv (c_{-1}^2+c_{+1}^2)/2$ and a 
phase $\alpha \equiv\arctan(c_{-1}/c_{+1})$. For a smearing angle of 
$15^\circ$ the LSS model predicts $d=0.0226$, $\alpha = 73^\circ$, which has 
to be compared to $d=0.0138 \pm 0.0049$, $\alpha = 89^\circ \pm 20^\circ$, 
obtained from the joint data set of~\cite{Aab:2014ila}.  Notice that 
the incompatibility with the LSS models worsens when we include the $z$ 
component to form $C_1$, despite the fact that the largest error comes with 
$a_{1,0}$; this is because the data value for $a_{1,0}$ is very small 
compared to that of the LSS model --- the latter is within a factor of 3 
from the other dipole coefficients.

The relations~(\ref{eq:c_+-}) become exact if the flux contains only a 
dipole and any other even multipole, but has zero power for all odd $\ell>1$; 
in this approximation these coefficients have also been measured by the 
Pierre Auger collaboration alone~\cite{Abreu:2011ve,Abreu:2012ybu} --- in 
this case the errors on these quantities are indeed smaller: the price to 
pay is a \emph{a priori} decision on what the flux should be.

\section{The GMF and the harmonic multipoles}
\label{sec:GMF}

The results of the previous section did not take into account the effects of
the propagation of UHECRs through the GMF directly, but only indirectly
through a variable, and relatively large, smearing angle. A better
approximation is to treat the regular part of GMF explicitely and leave the
smearing to represent the random deflections only, deflections which can 
amount to about $10^\circ$ to $20^\circ$ for our 10 EeV protons, see~\cite{Pshirkov:2013wka,Beck:2014pma}. So one may 
wonder whether the regular GMF could wipe away, or distort, any anisotropic 
harmonic imprint, for example by transferring power from a multipole to 
another.  A caveat here is that the regular GMF is not known well enough for
accurate predictions of $a_{\ell m}$. 

We will show, however, that while the direction-dependent $a_{\ell m}$ are
indeed quite sensitive to the strength (and shape) of the magnetic field in
the Milky Way, the direction-blind power spectrum $C_\ell$ is quite stable
against these perturbations.

In order to demonstrate this empirically we adopted the model of~\cite{Pshirkov:2011um} for the regular GMF, and we simulated again the
expected fluxes from the LSS, now propagating the flux through the
GMF. This GMF model has two components, a disk field and a halo
field, with independent strengths.  We chose to work with the
best-fit parameters as reported in~\cite{Pshirkov:2011um} for the version dubbed
bisymmetric spiral structure, or BSS: this means that the overall disk
and halo strengths are $B_\text{disk}=2\;\mu$G and
$B_\text{halo}=4\;\mu$G, respectively. We should stress, however, that neither
the choice of the model parameters, nor the model itself has too strong an impact
on our results, as what we found is that the effect of GMF on the coefficients
of the power spectrum is small.

\begin{figure}
\begin{center}
  \includegraphics[width=0.48\textwidth]{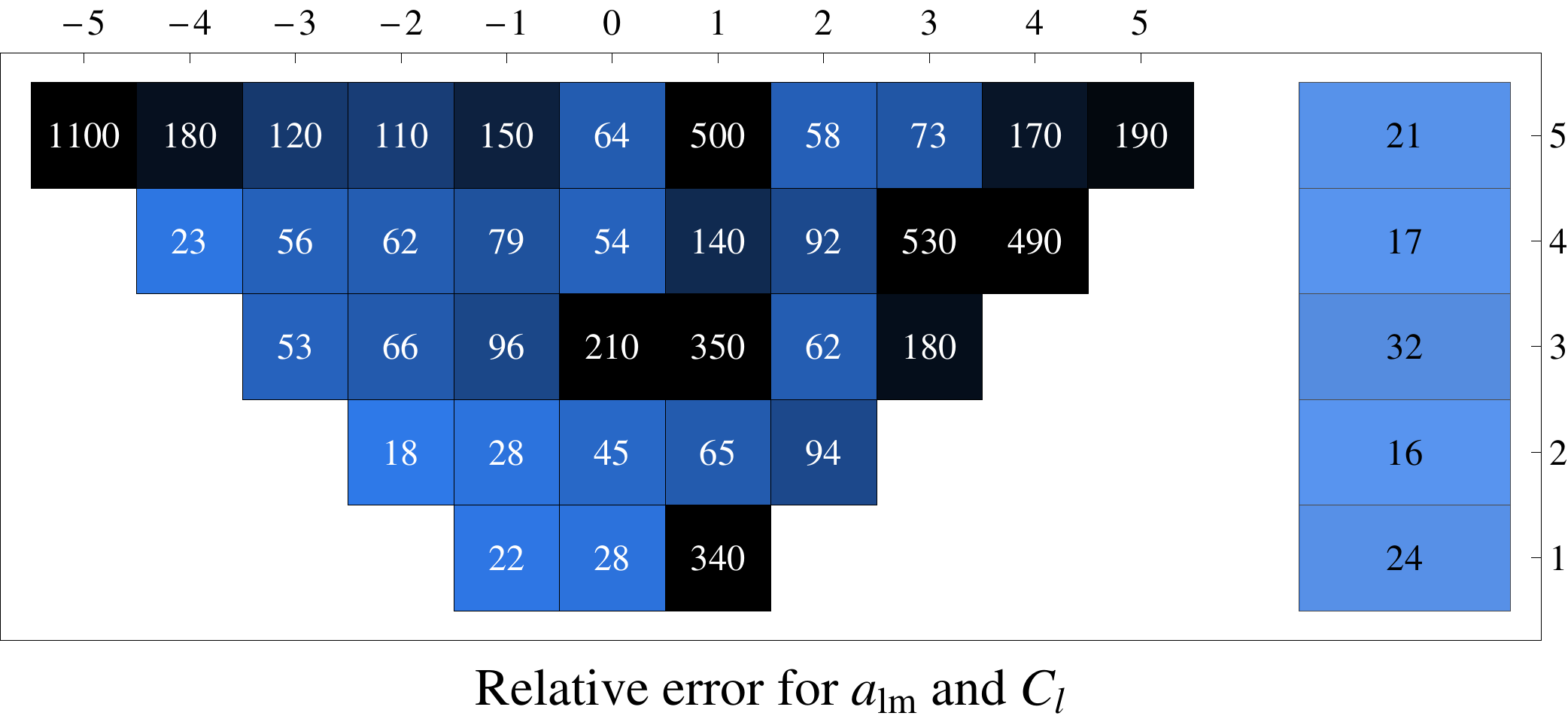}
\end{center}
\caption{Colour-coded relative variation of the $a_{\ell m}$ and the $C_\ell$ 
with varying magnetic field strength (in percent).}
\label{fig:pyr}
\end{figure}

To assess the variability, or sensitivity, of the harmonic coefficients and
power spectrum on the strength of the GMF --- a similar virtual experiment can
be performed by changing its shape, or both --- we generated 1000
fluxes\footnote{When propagating UHECRs through the magnetic field we use
  monochromatic primaries, since the deviations are maximised at the lowest
  energy, for simplicity.} with randomly chosen field strengths ranging from
zero to twice the best fit values, that is $B_\text{disk}=[0,\,4]\,\mu$G and
$B_\text{halo}=[0,\,8]\,\mu$G.  Since each time the reconstructed flux will be
slightly different, we show the relative percent variation (standard deviations over the mean: $\sigma_x/\text{mean}_x$, where $x$ stands for the $a_{\ell
  m}$ and $C_\ell$, in Fig.~\ref{fig:pyr}.  We immediately notice that the
average spread for some harmonic coefficients $a_{\ell m}$ is much larger than
that of the power spectrum coefficients $C_\ell$, proving our point above.  We 
show in the picture only multipoles up to $\ell=5$, but we performed the same 
exercise for multipoles up to $\ell=20$, and obtained the same result. 

A legitimate doubt is that since we do not expect these distributions to be
Gaussian, as we are not sampling the same sky many times with different
randomly generated events, a very skewed distribution may bias this result and
the standard deviations would not represent the actual excursion of the
quantities under observation.  For example, the spread of the $C_\ell$ might
be a bad indicator of how much the power spectrum actually varies, but be
accurate in describing the fluctuations of $a_{\ell m}$. We have checked that
the ratio between the total excursion for a given parameter, that is, its
maximum minus its minimum, versus the corresponding deviation,
$|\max(x)-\min(x)|/\sigma_x$, is more or less the same (to within 15\%) for all the quantities
we analyse, which means that for both the $a_{\ell m}$ and the $C_\ell$
coefficients, the spread is an equally good descriptor of the range which
these parameters can attain.

The conclusion we draw from these tests is that:
\begin{itemize}
  \item the power spectrum is a much more suitable quantity in
    assessing the anisotropic properties of the UHECR flux when
    dealing with the GMF, as it is much more robust against the, still
    poorly known, details of the GMf itself, compared to the
    direction-dependent $a_{\ell m}$;
  \item the absolute power spectrum itself is not much affected
    quantitatively by the regular GMF: we thus believe it is unlikely
    that the reason behind the low quadrupole observed in the data is
    to be found in the effect of GMF UHECRs deflections.
\end{itemize}

\section{Isotropic fraction and the harmonic multipoles}
\label{sec:Fe}

So far we have always worked with proton primaries, but the UHECRs
composition at the highest energies is not known.  If instead of
protons we were to propagate iron nuclei, the deflections they endure would 
be a factor 26 larger, due to the corresponding larger rigidity.  We
thence expect that a fraction of iron or other nuclei in the total
UHECR flux, because of its tendency to isotropise, would help
loosening the tension between the observed and simulated multipoles.

To assess this, we again generated several flux maps where we subtract a
fraction of the total proton flux and replace it with an isotropic one, to
roughly simulate the contribution of iron.  We vary this ``iron fraction''
(essentially, the isotropic fraction) between zero and one, and recalculate
the power spectrum for each map; we then compare with the data and their
errors, and compute the statistical significance of the low $\ell$ coefficients of
the power spectrum in each map.  Were the primaries a mix of several different elements, the LSS predictions would fall in between the values we obtain below.

\begin{figure}
\begin{center}
  \includegraphics[width=0.42\textwidth]{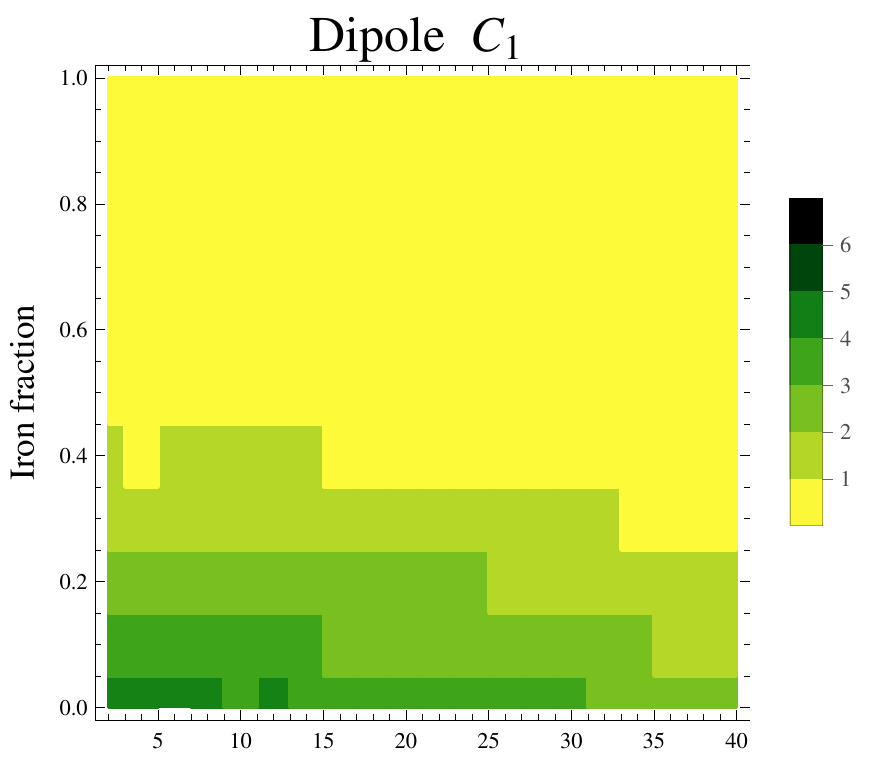}\\\vspace{10pt}
  \includegraphics[width=0.42\textwidth]{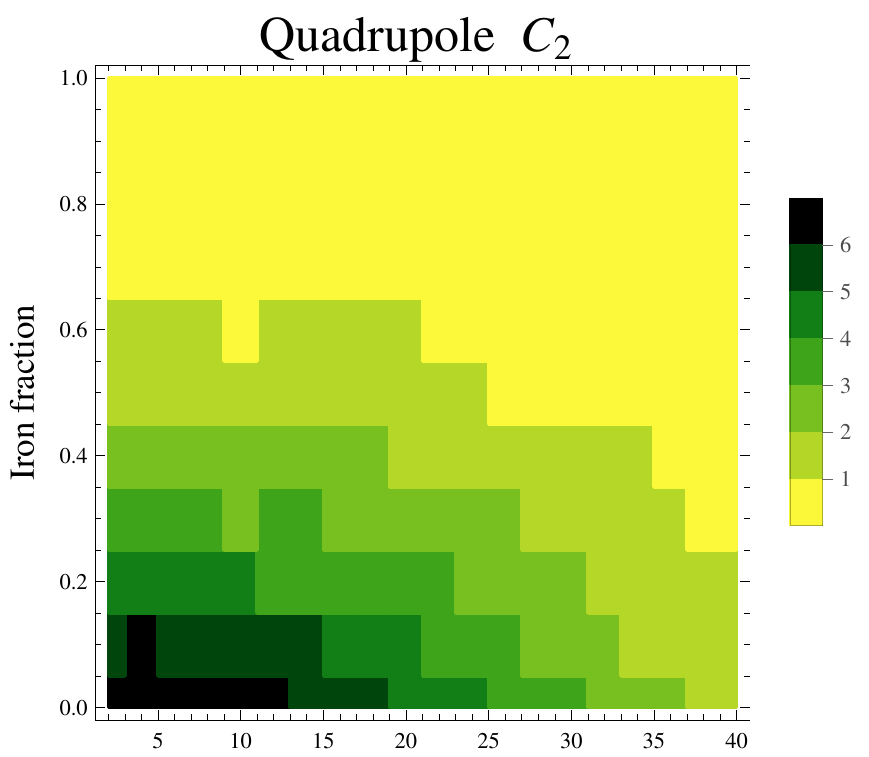}\\\vspace{10pt}
  \includegraphics[width=0.42\textwidth]{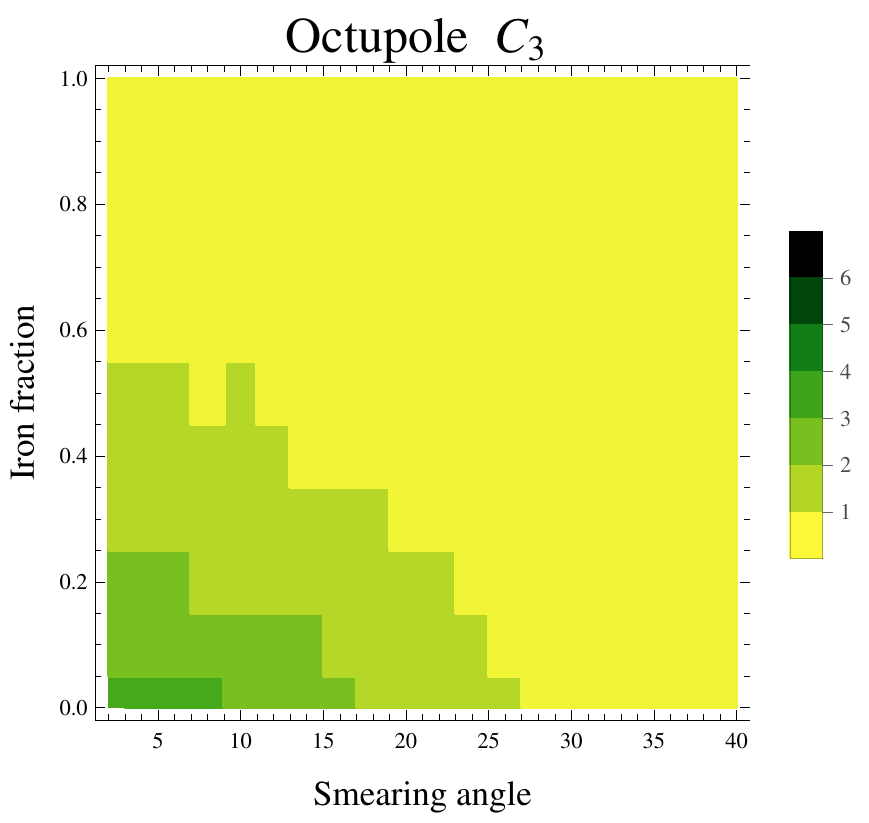}
\end{center}
\caption{Dipole, quadrupole, and octupole against LSS expectations for
  varying iron fraction and turbulent GMF -- no regular GMF.}
\label{fig:rigoff}
\end{figure}

In Fig.~\ref{fig:rigoff} we show $1\,\sigma$, $2\,\sigma$, $3\,\sigma$,
etc, contours for the dipole $C_1$, quadrupole $C_2$, and octupole
$C_3$, where in addition to varying the isotropic fraction, we also change
the smearing angle of the map, to account for a variable turbulent GMF
strength.

\begin{figure}
\begin{center}
  \includegraphics[width=0.42\textwidth]{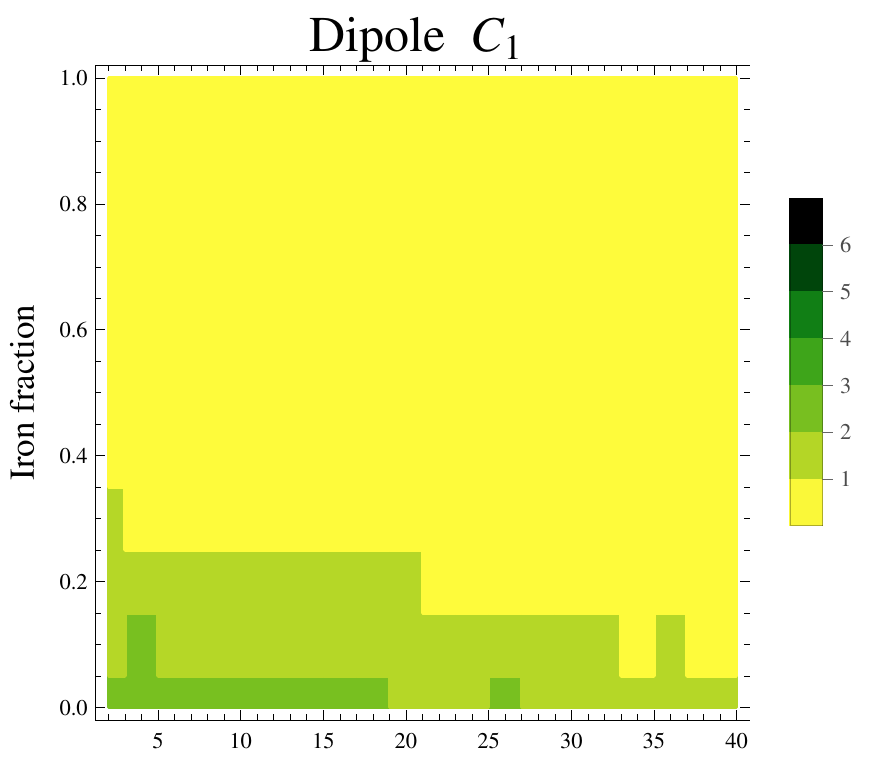}\\\vspace{10pt}
  \includegraphics[width=0.42\textwidth]{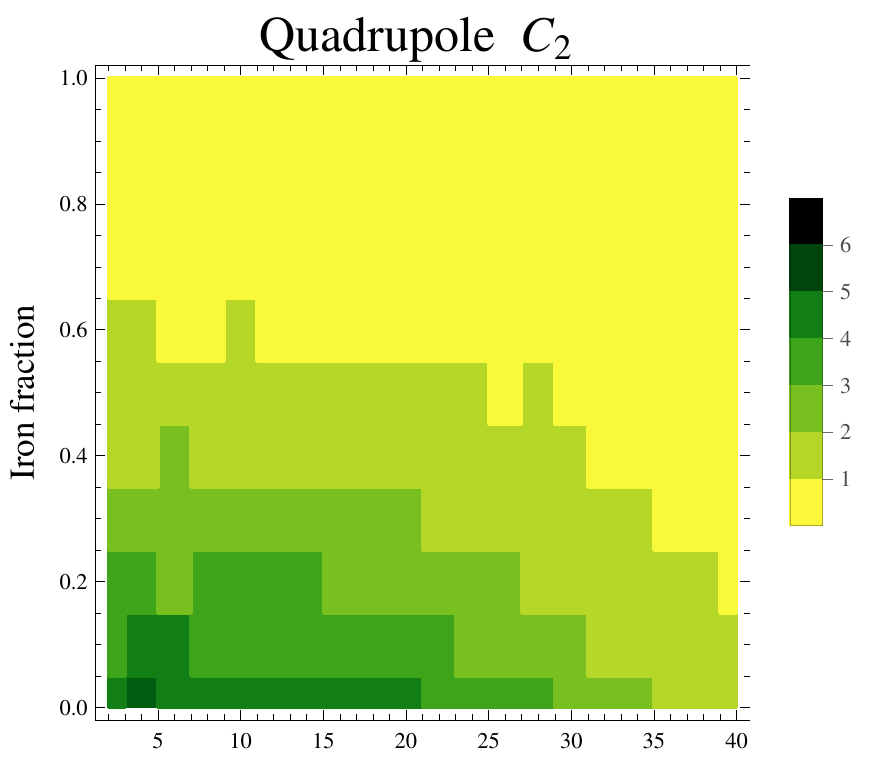}\\\vspace{10pt}
  \includegraphics[width=0.42\textwidth]{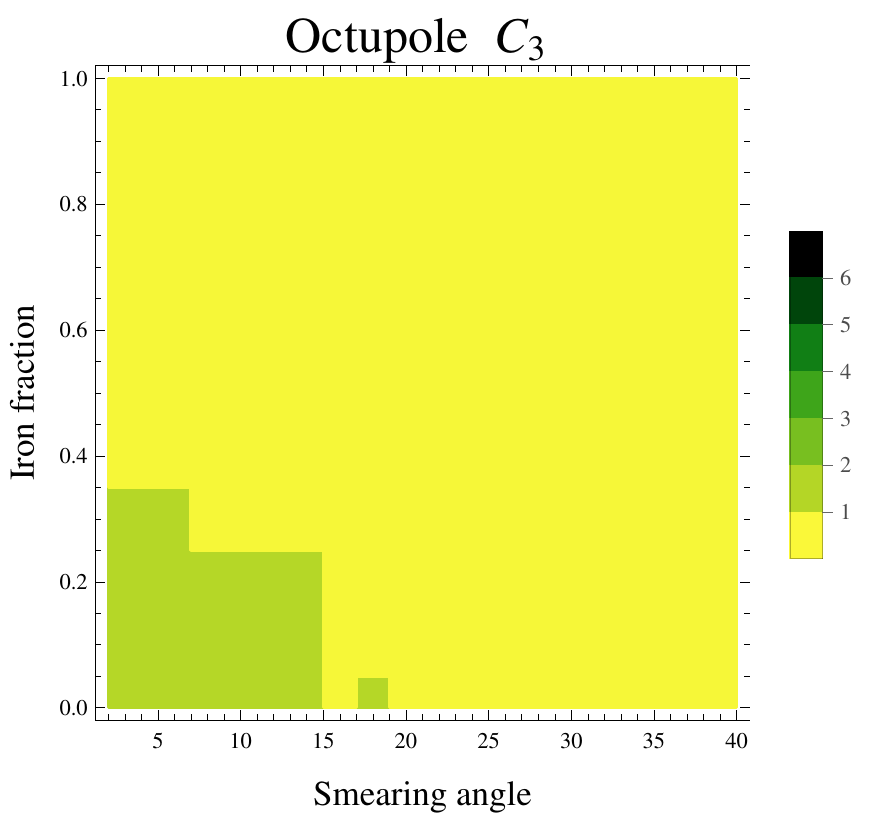}
\end{center}
\caption{Dipole, quadrupole, and octupole against LSS expectations for
  varying iron fraction and turbulent GMF -- with regular GMF.}
\label{fig:rigon}
\end{figure}

We can perform this test with or without the regular GMF, and the results,
according to our previous section, should not change much; this is indeed the
case, as Fig.~\ref{fig:rigon} shows: the curves move down by approximately
$1\sigma$, which again does not suffice to resolve the tension between data
and LSS expectations.

As we see, both the dipole and the quadrupole in the data prefer a more
isotropic Universe, with the quadrupole being the most pronouncedly
incompatible with the expectations from LSS.

Correlating the arrival directions of UHECRs with LSS is a logical surmise, 
so this result is somewhat puzzling; however, it is not completely 
unexpected, as at this energy it is known that, for example, the TA data 
alone prefer isotropy to LSS~\cite{AbuZayyad:2012hv}.  What is shown here is then 
simply another parametrisation of the same result, but one which can bring 
some insight into the physics behind it. For instance, it may be that there 
is a bias, or systematic effect which causes the dipole and/or the 
quadrupole to have a surprisingly low power --- such an effect could arise due to an excess on the galactic plane, for example.

It would be extremely interesting to be able to look at the same figures
above, say, 60 EeV, where instead an isotropic flux is incompatible with TA
data at more than $3\sigma$ at not too large smearing angles~\cite{AbuZayyad:2012hv}; the same data
are more in line with the predictions from LSS.  Unfortunately, the very low
statistics in the common band of the two experiments (and consequently, large
errors) makes this task quite futile at present.

Finally, as expected, the random GMF, mimicked through the smearing angle,
does not affect the low-$\ell$ part of the power spectrum significantly ---
only at large smearing angles features tend to be blurred, so the differences
between the LSS flux and an isotropic one are diluted.

\section{Conclusion}
\label{sec:end}

For the first time we have a full sky map of UHECRs; this opens up the possibility to decompose the flux on the sphere in a harmonic basis, and obtain its angular power spectrum $C_\ell$, which is shown in Fig.~\ref{fig:lss}.  We wanted to see whether, in this language, a source distribution which traces that of matter (galaxies), would produce the same $C_\ell$.

We find that, assuming lone proton primaries, and discarding for now the GMF, this is not the case.  In particular, LSS models tend to generate a much larger power that what is extracted from the data, especially at low multipoles such as the dipole and quadrupole; the experimental full sky map is much more isotropic.  The discrepancy for the quadrupole $C_2$ can be as strong as about $6\sigma$, while in the case of the dipole $C_1$ at small smearing angles the data value is $4\sigma$ away from the LSS prediction for it.

When we turn on the regular GMF we observe a strong correlation between the variability of the $a_{\ell m}$ values and the strength and shape of the GMF; at the same time, the power spectrum $C_\ell$ is much more stable against the same perturbations.  The latter is therefore a more reliable observable in investigations like the one we present here.  This also means that the incompatibility between data and LSS is \emph{not} an artifact of ignoring the GMF.

Since the data prefers a more isotropic Universe, one possibility is that the primaries are heavy, and diffuse in the GMF (both regular and random).  We introduced heavy nuclei in our study in the guise of a, variable, isotropic fraction of the total CRs flux to understand how much more isotropic the distribution of UHECRs sources needs to be: in some cases (quadrupole at small and intermediate smearing angles) to tame the LSS prediction the one may need up to about 50\% or more of isotropic flux fraction.

In fact, since our method simply discriminates between an anisotropic proton flux and an isotropic one, alternative explanations are possible, not related to the composition of the promaries.  An additional isotropic component may be the result of acceleration mechanisms operative away from galaxies; alternatively, the more isotropic distant sources may be contributing more than expected (for instance due to exotic particle interactions); one more possibility is that our Galaxy is plunged into a strong magnetic wind, which isotropises the arrival directions of UHECRs even before they reach the Milky Way.

At 10 EeV, the energy threshold of the datasets we used in this work, the outcome of our analysis are not a surprise, as the data are known to be incompatible with LSS models; the multipolar description of the same result (but now with data from the full sky) can help in identifying the physical reason behind it: for instance, the $C_1$ and $C_2$ results may be signalling the presence of some systematic effect.  On the other hand, the 60 EeV data does prefer LSS: it would be extremely useful to be able to repeat our exercise at those energies, but with current data this would be inconclusive.  In the future, the source of these discrepancies could be identified, and it will bring some crucial insight into the hunt for UHECRs sources.


\section*{Acknowledgements}

This paper has been prepared for a special volume dedicated to V.~Rubakov's 60th anniversary.  We seize this opportunity to wish him --- a teacher and an old friend for PT, a colleague for FU --- many more fruitful and happy years.

FU and PT are supported by IISN project No.~4.4502.13 and Belgian Science
Policy under IAP VII/37. PT is supported in part by the RFBR grant
13-02-12175-ofi-m.

\section*{Appendix}
\label{sec:app}

The flux distribution on the sphere $\Phi(\bn)$ can be projected on the right
ascension plane and expanded in a Fourier series on the circle:
\[
  \frac12 \int\text{d}\cos{\theta} \Phi(\theta,\phi) \equiv \sum_n c_n Y_n(\phi) \, ,
\]
whence
\[
  c_n = \frac12 \int\text{d}\bn \Phi(\bn) Y_n(\phi) \, .
\]
In order to derive the relation between the projected $c_n$ and the spherical harmonics coefficients $a_{\ell m}$ we make use of Eq.(\ref{eq:ylm}) and obtain
\[
  c_n = \frac{2^{|n|}|n|}{8} \sum_\ell \left[ (-1)^{|n|} 
+ (-1)^\ell \right] a_{\ell n} 
\sqrt{\frac{2\ell+1}{2}\,\frac{(\ell-|n|)!}{(\ell+|n|)!}} 
\]
\[
\times  \frac{\Gamma[\ell/2]\,\Gamma[(\ell+|n|+1)/2]}{\Gamma[(\ell+3)/2]\,\Gamma[(\ell-|n|+2)/2]} \, .
\]
This implies that, \emph{if and only if} $\Phi(\bn)$ is a pure dipole plus any other $\ell\in\text{even}$, but does not contain any $\ell\in\text{odd}$:
\[
  c_{\pm1} = \frac{\sqrt3\pi}{8} a_{1,\pm1} \, ,
\]
from which $\alpha = \arctan(c_{-1}/c_{+1}) = \arctan(a_{1,-1}/a_{1,1})$ follows.  Since in reality there will be more multipoles, this is only a first approximation to the actual result.  We have checked, again up to $\ell=20$, that in the case of our LSS models, at the level of precision we can attain with the common analysis the higher multipoles contribute below the errors, and they become, as they should, progressively unimportant.


\bibliographystyle{unsrt}
\bibliography{LowQ_bib}


\end{document}